\documentclass{PoS}

\title{A jet model for black-hole X-ray sources} 

\ShortTitle{Jet model}

\author{\speaker{N. D. Kylafis}, I. E. Papadakis, P. Reig\\
        \\
	University of Crete and\\
	Foundation for Research and Technology - Hellas, Greece\\
        \email{kylafis@physics.uoc.gr}}

\abstract{A jet model for Galactic black-hole X-ray binaries will be 
presented that appears to explain several observational
characteristics.  In particular, it explains the energy spectrum from 
radio to hard X-rays, the time-lags as a function of Fourier frequency,
the increase of the variability amplitude (QPO and high frequency) 
with increasing photon energy, and the narrowing of the autocorrelation
function with increasing photon energy.  On the other hand, there are 
additional observational constraints that no model has tried to explain
yet.  It is important that we all try to address these constraints 
if we are to make any progress in understanding black-hole X-ray sources.
}

\FullConference{VI Microquasar Workshop: Microquasars and Beyond\\
		September 18-22 2006\\
		Societ\`a del Casino, Como, Italy}

\begin{document}

\section{Introduction}
We will be a little provocative in our contribution, not 
because we think that we have the right answers to outstanding questions
regarding black-hole X-ray binaries (or even some of the right answers),
but because we want to bring forward and stress some important 
observational conclusions (new and old) that most models do not address.

\subsection{Observational conclusion 1}
Miller et al. (2006) found that an observation of GX 339-4 revealed 
an accretion disk extending to the Inner Stable Circular Orbit (ISCO), 
when the source was in the low/hard state.  The same is true for 
SWIFT J1753.5-0127 (Miller, Homan, and Miniutti 2006).
We understand that the above conclusion is not universally accepted, 
as the common belief is that the accretion disk in the low/hard state
is truncated at some inner radius larger than the ISCO.  Nevertheless,
it is accepted by the community that the data allow for such a 
conclusion.  

If this conclusion is correct, then the radio jet that is seen in the 
low/hard state is not associated with a truncated accretion disk.  
This then, leads us to suggest that there is no reason to exclude a jet 
(weak as it may be) when the source is in the high/soft state and the
accretion disk extends all the way in to the ISCO.

Even if the above conclusion is proven to be incorrect, we still
suggest that a weak (undetected up to now) jet may be present when the
black-hole X-ray source is in the high/soft state.  We will support this
idea below.

\subsection{Observational conclusion 2}
Fourier resolved spectroscopy has shown that the accretion disk is 
not variable, at least in three black-hole X-ray sources:  
Cyg X-1 (Revnivtsev, Gilfanov, and Churazov  1999; 
Gilfanov, Churazov, and Revnivtsev  2000; 
Churasov, Gilfanov, and Revnivtsev  2001), 
GX 339-4 (Revnivtsev, Gilfanov, and Churazov 2001), 
and 4U 1543-47 (Reig et al. 2006).

It is inescapable then, that the observed variability in hard X-rays
comes from the jet (or whatever you want to call that thing above 
and below the accretion disk, e.g., corona, wind, flares, etc.).

\subsection{Observational conclusion 3}
It has been shown by Pottschmidt et al. (2000) that the time-lags as a
function of Fourier frequency are identical in the low/hard and the 
high/soft states.  Thus, it is natural to assume that identical
mechanisms must be producing the time-lags in both the low/hard and
the high/soft states.  This introduces significant constraints to 
models.  In addition, if the jet introduces the time-lags in the 
low/hard state, then a weak jet must be present in the high/soft 
state.

\section{Jet model}
For whatever it is worth, the jet model proposed by Reig, Kylafis, 
and Giannios (2003) explains the time-lags in a natural way as time 
delays due to upscattering of soft photons in the jet.

To understand how the time-lags can be inversely proportional to
Fourier frequency, consider the following simplified picture of
the jet (Fig. 1).  Let the jet consist of a series of spheres with
increasing radii $R_i$.  Arrange the densities of the spheres so that a
soft photon from the source below (denoted by a * in the Figure)
has equal probability to scatter in any of the spheres.  Let the 
time delay due to scattering in the sphere with radius $R_i$ be $t_i$.
It is well known that Compton scattering that causes a delay $t_i$
acts as a filter and cuts off all frequencies higher that $\sim 1/t_i$.
In other words, the larger the Fourier frequency that is observed
the smaller the time-lag.

The model of Reig et al. (2003) 
makes the assumption that the density in the jet varies 
inversely proportional with height.  This ensures that the optical 
depth to electron scattering is the same per decade of height.  Thus,
if the total optical depth along the jet is of order unity, the soft
input photons scatter with equal probability at all heights in the jet.
Thus, all time lags from zero to a maximum value are sampled by the
soft photons, but only frequencies less than the inverse time lag 
survive and are observed.
A Monte Carlo calculation (Reig et al. 2003) has verified that the time
lags are inversely proportional to Fourier frequency.

Two additional constraints regarding the flattening of the power 
spectrum  at high frequencies and the narrowing of the autocorrelation 
function with increasing photon energy has also been explained with 
the jet model (Giannios, Kylafis, and Psaltis 2004).

Also, for XTE J 1118+480, the only source for which we have simultaneous
observations from radio to hard X-rays, the energy spectrum is reproduced 
extremely well over 10 orders of magnitude in energy by the jet model 
(Giannios 2005).  It must be stressed here, that the energy spectra 
and the time variability are explained with the same values of the 
parameters of the model.

\begin{figure}
\begin{center}
\includegraphics[width=.6\textwidth]{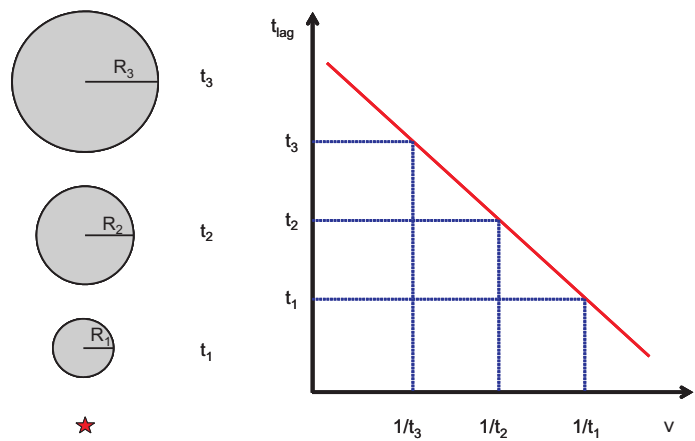}
\caption{Schematic of a "jet" with properties discussed in the text.}
\label{fig1}
\end{center}
\end{figure}

\section{What next?}
We are of the opinion that hardly any model in Astrophysics is entirely
correct!  Thus, we think that all models should be pushed to their limits.
Some recent observational facts place very stringent constraints on all
models.

Pottschmidt et al. (2003) have fitted the power spectra of Cyg X-1 
(over a period of \~ four years) with four broad Lorentzians with 
peak frequencies $\nu_1$, $\nu_2$, $\nu_3$, and $\nu_4$.

\subsection{Unexplained observational fact 1}
The ratios of the peak frequencies of the above four Lorentzians are 
constant with time (see Fig. 5 of Pottschmidt et al. 2003).  
This means that the four Lorentzians "talk" to each other.

\subsection{Unexplained observational fact 2}
In Fig. 7a of Pottschmidt et al. (2003), one can see that the average 
time lags (in the 3.2 - 10 Hz range) are correlated with the peak
frequencies $\nu_1$, $\nu_2$, and $\nu_3$.  

Similar correlations (Fig. 7b) have been found between the power-law spectral 
index $\Gamma$ and the peak frequencies $\nu_1$, $\nu_2$, and $\nu_3$.

It is not surprising then that $\Gamma$ is correlated (Fig. 7c) with the
average time lags (in the 3.2 - 10 Hz range).

\section{Summary}
In our opinion, the jet in black-hole X-ray binaries is not simply an
ornament.  It appears to be responsible not only for a large part of the 
energy spectrum (radio, optical, hard X-rays), but also for the
observed time variability.  

The three broad Lorentzians in the power spectrum with peak frequencies
$\nu_1$, $\nu_2$, and $\nu_3$ seem to be produced by one mechanism 
(since their ratios are constant) and they "communicate" with the
power-law spectral index $\Gamma$.  In other words, each observed
value of $\Gamma$ in Cyg X-1 is associated with specific values of the
peak frequerncies $\nu_1$, $\nu_2$, and $\nu_3$.  

With such stringent constraints, life will be difficult for modelers, 
but no progress in our understanding will be made unless we address
all the observational facts simultaneously.  Addressing only the spectra
or only the time lags or only some observational facts will not lead 
us anywhere.

{\it Acknowledgments:} This work has been supported in part by a 
PYTHAGORAS II grant from the Ministry of Education of Greece.

\end{document}